\documentclass[pre,aps,twocolumn,showpacs,amsmath,amssymb,amsfonts,floatfix,superscriptaddress]{revtex4}
\usepackage{graphicx}
\usepackage{dcolumn}
\usepackage{bm}
\usepackage{latexsym}
\usepackage{float}
\usepackage{supertabular}
\usepackage{longtable}
\usepackage{verbatim}
\usepackage{hyperref}
\usepackage{color}
\usepackage{subfigure}

\begin{document}
\title{Influence of Luddism on innovation diffusion}
\author{Andrew Mellor}
\affiliation{Department of Applied Mathematics, School of Mathematics, University of Leeds, Leeds LS2 9JT, U.K.} %
\author{Mauro Mobilia}
\affiliation{Department of Applied Mathematics, School of Mathematics, University of Leeds, Leeds LS2 9JT, U.K.} %
\author{S. Redner}
\affiliation{Santa Fe Institute,1399 Hyde Park Rd., Santa Fe NM, 87501, U.S.A} %
\affiliation{Center for Polymer Studies and Department of Physics, Boston University, Boston,
MA 02215, U.S.A.}
\author{Alastair M. Rucklidge}
\affiliation{Department of Applied Mathematics, School of Mathematics, University of Leeds, Leeds LS2 9JT, U.K.} %
\author{Jonathan A. Ward}
\affiliation{Department of Applied Mathematics, School of Mathematics, University of Leeds, Leeds LS2 9JT, U.K.} %

\begin{abstract}
  We generalize the classical Bass model of innovation diffusion to include a
  new class of agents---Luddites---that oppose the spread of innovation.  Our
  model also incorporates ignorants, susceptibles, and adopters.  When an
  ignorant and a susceptible meet, the former is converted to a susceptible
  at a given rate, while a susceptible spontaneously adopts the innovation at
  a constant rate.  In response to the \emph{rate} of adoption, an ignorant
  may become a Luddite and permanently reject the innovation.  Instead of
  reaching complete adoption, the final state generally consists of a
  population of Luddites, ignorants, and adopters.  The evolution of this
  system is investigated analytically and by stochastic simulations.  We
  determine the stationary distribution of adopters, the time needed to reach
  the final state, and the influence of the network topology on the
  innovation spread.  Our model exhibits an important dichotomy: when the
  rate of adoption is low, an innovation spreads slowly but widely; in
  contrast, when the adoption rate is high, the innovation spreads rapidly
  but the extent of the adoption is severely limited by Luddites.
\end{abstract}

\pacs{05.40.-a, 02.50.-r, 89.75.-k, 89.65.-s}

\maketitle

\section{Introduction}

Models of innovation diffusion seek to understand how new ideas, products,
or practices spread within a society through various channels~\cite{R03}.
Innovation may refer to new technologies or deviations from existing social
norms.  Rather than a single theory, innovation diffusion represents a
theoretical framework that encompasses a range of social models in which the
term ``diffusion" can mean contagion, imitation, and social
learning~\cite{Kincaid04,R&G43,CKM}.

Many of the traditional approaches~\cite{1960s} to innovation diffusion
modeling are based on a mean-field approximation and are referred to as
aggregate models.  An influential example is the seminal Bass
model~\cite{B69,B80,M90,B04,CEJOR2012,Hopp04,BMapplications}, where
innovation spreads as the result of either an adopter converting a
susceptible (contagion), or through external influences on susceptibles
(advertising and mass media).  The basic outcome of the Bass model is that the
time dependence of the fraction of adopters exhibits a sigmoidal
shape~\cite{B69,B80,M90,B04,R03,PY09}.  Thus significant adoption arises
only after some latency period, after which complete adoption is quickly
achieved.

While the Bass and related models have been successful in fitting historic
data~\cite{Sultan90}, there are several limitations of these approaches:
\begin{itemize}
\item The predictive power  of the Bass
  model is uncertain~\cite{BMparameters,Hohnisch08}.

\item Aggregate models are based on infinitely large, homogeneous
  populations~\cite{CEJOR2012,PY09} and  cannot account for
  sample-specific differences and related fluctuation phenomena.

\item Bass-like models do not account for behavioral patterns that result from
  social reinforcement and ``bandwagon'' pressure~\cite{C10,Cth,bandWagon,reinf}.

\item Aggregate models assume a ``pro-innovation bias'' and thus cannot
  reproduce phenomena such as incomplete
  adoption~\cite{G00,R03,C10,bandWagon}.
\end{itemize}

We are particularly interested in situations where innovation can be
accompanied by controversy, suspicion, or rejection within some social
circles, potentially leading to incomplete adoption.  As an example, mobile
phones are owned by 90\% of Americans~\cite{mobiles1} as of 2014, but their
use is accompanied by continued health and safety concerns~\cite{mobiles2}.
Similarly, the coverage of the measles, mumps and rubella vaccine in the
United Kingdom reached 92.7\% in 2013--14, below the target level of 95\%
coverage for herd immunity.  This incomplete adoption level may result from
doubts about vaccine effectiveness and safety concerns promulgated by
anti-vaccination movements~\cite{vaccine1,vaccine2}.  Such doubts seem to
persist even in the face of their apparently negative consequences, such as
the measles epidemic that seemed to have its inception in Disneyland at the
start of 2015.

Motivated by these facts, we introduce a model for the diffusion of an
innovation, using a statistical physics approach~\cite{DOInets}, in which we
account for the competing role of ``Luddites'' in hindering the spread of the
innovation.  Agents may either be Luddites (opposed to innovation),
``Ignorants'' (no knowledge of the innovation), ``Susceptibles'' (receptive
to innovation), or ``Adopters'' of the innovation.  We dub this the LISA
(Luddites/Ignorants/Susceptibles/Adopters) model.  The main new feature of
the LISA model is the existence of agents that reject the innovation in
response to the spread of adoption. 
Previous work \cite{resistance} has considered the introduction of `resistance leaders' who spread a negative response to the innovation, akin to the spread of a competing innovation. The LISA model differs from this approach by considering agents who respond in particular to the rate of uptake of the innovation.
We use the term ``Luddites'' in
reference to the 19th-century Luddism movement in which English textile
artisans protested against newly developed labor-saving
machinery~\cite{Luddites_hist}.  We are interested in determining how Luddism
limits the final level of adoption and how the presence of Luddites leads to
a trade-off between adoption levels and adoption times scales.

The LISA model is defined in the next section, while the behavior of the
model in the mean-field limit and on complete graphs is investigated in
Section~\ref{sec:MF}.  Section~\ref{sec:graphs} focuses on the model dynamics
on random graphs and on a one-dimensional regular lattice.  For all these
substrates, we investigate how Luddism affects the final level of adoption
and the time scale of adoption.  We also elucidate a dichotomy between the
cases of slow but relatively universal adoption for low values of an
intrinsic innovation rate, and the rapid but limited spread of innovation
that occurs in the opposite limit.  Our conclusions are presented in
Section~\ref{sec:conc}.

\section{The LISA model}

As a helpful preliminary, let us review the simpler two-state Bass model of
innovation diffusion.  Here a population consists of two types of agents:
susceptibles $\mathcal{S}$ or adopters $\mathcal{A}$.  In the Bass model,
susceptibles can become adopters via either of two processes:
\begin{itemize}
\item[(a)] Contagion-driven conversion: a susceptible converts to an adopter
  by interacting with another adopter, as represented by the process
  $\mathcal{S}+\mathcal{A} \to \mathcal{A}+\mathcal{A}$.
\item[(b)] Spontaneous adoption: a susceptible converts to an adopter,
  $\mathcal{S} \to \mathcal{A}$.
\end{itemize}
The characteristic feature of the Bass model is that the adopter density exhibits a 
sigmoidal time dependence, in
which the time derivative of this density has a sharp peak (corresponding to
an inflection point in the time dependence of the density itself), before
complete adoption eventually occurs~\cite{B69,B80,M90,B04,R03,PY09}.

\begin{figure}[ht]
\centering
\includegraphics[width=0.8\linewidth]{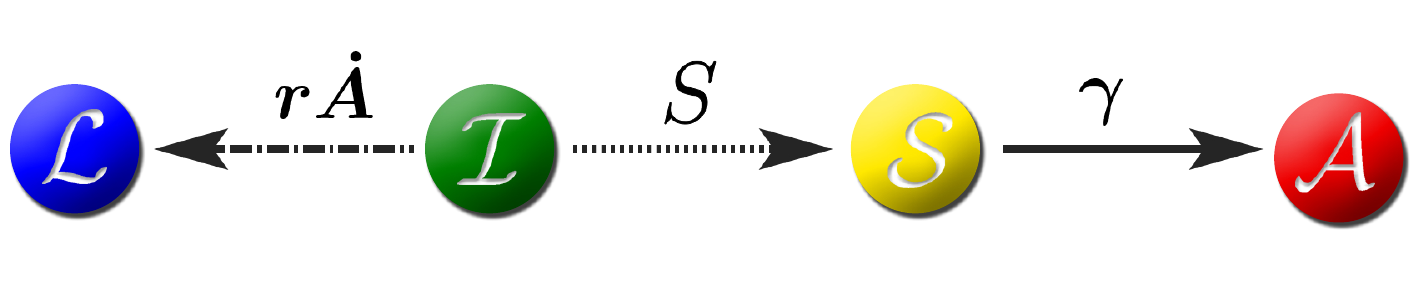}
\caption{(\textit{Color online}) Schematic depiction of the LISA model.  An
  ignorant $\mathcal{I}$ can become a Luddite $\mathcal{L}$ with rate
  $r\dot{A}$ (in a mean-field setting); an ignorant can also become a
  susceptible $\mathcal{S}$ by contagion with rate proportional to the
  susceptible density.  A susceptible spontaneously becomes an adopter at
  rate $\gamma$.}
\label{fig:model}
\end{figure}

Our LISA model is a four-state system that consists of a population of $N$
individuals that can each be in the states of Luddite ($\mathcal{L}$),
ignorant ($\mathcal{I}$), susceptible ($\mathcal{S}$), or adopter
($\mathcal{A}$).  Ignorant agents may either be persuaded to become
susceptible, and thence reach the adopter state, or they may become a Luddite
and permanently oppose the spread of the innovation.  Specifically, the
elemental steps of our LISA model are the following (see
Fig.~\ref{fig:model}):
\begin{enumerate}
\item[(a)] Contagion-driven conversion: An ignorant agent becomes susceptible
  by interacting with another susceptible agent.  That is,
  $\mathcal{I}+\mathcal{S} \to \mathcal{S}+\mathcal{S}$ with rate 1.

\item[(b)] Spontaneous adoption: A susceptible agent spontaneously becomes an
  adopter, $\mathcal{S} \to \mathcal{A}$ with rate $\gamma$
\footnote{Adoption could also occur by contagion, according to
  $\mathcal{S} + \mathcal{A} \to \mathcal{A} + \mathcal{A}$. Yet, 
this two-body process would yield  similar features as our LISA model, 
but would be technically more tedious to handle.}.

\item[(c)] Luddism: Ignorants may permanently reject the innovation
  and become Luddites, $\mathcal{I}\to \mathcal{L}$, with a rate
  proportional to the change in the density of adopters in its
  neighborhood.
\end{enumerate}

The Luddism mechanism outlined above incorporates two aspects of
negative behavior towards innovation.  The first represents a fear of
innovation or its consequences, as in the case of the historical
Luddism movement, where the introduction of labor-saving machinery
caused fear over job security~\cite{Luddites_hist}.  The second is
that of non-conformity; agents may oppose the innovation simply due to
its rapid increase in popularity~\cite{bandWagon}.  We model this
feature by defining the rate at which the Luddite density increases to
be \emph{proportional} to the adoption rate, with constant of
proportionality denoted by $r$, the Luddism parameter.

The multi-stage progression $\mathcal{I} \to \mathcal{S} \to \mathcal{A}$ may
also be viewed as a type of social reinforcement mechanism in which
adoption follows from a succession of prompts from
neighbors~\cite{Cth,reinf}. 
The equivalent 3-state model with only Luddites, ignorants and adopters creates a polarized community creates a polarized community where the ratio of adopters to Luddites is dependent only on the Luddism parameter, $r$. Other relevant models \cite{resistance} have found that high levels of advertising can prompt a negative response to innovation which cannot be replicated with only three states.
The combination of a multi-stage progression
to adoption, together with the Luddite mechanism, arguably represents the simplest
generalization of the Bass model that gives rise to non-trivial long-time
state with incomplete adoption of an innovation. 

\section{Mean-field descriptions}
\label{sec:MF}

We first consider the LISA model in the mean-field limit, where agents
are perfectly mixed.  The densities of each type of agent are given by
$(L,I,S,A)=(N_{\mathcal{L}}, N_{\mathcal{I}}, N_{\mathcal{S}},
N_{\mathcal{A}})/N$, where $N_{X}$ is the number of agents
of type $X \in \{\mathcal{L}, \mathcal{I}, \mathcal{S},
\mathcal{A}\}$, and $N$ is the total number of agents.  We consider
the limit $N\to\infty$, so that all densities are continuous
variables and all fluctuations are negligible.  In this setting, the
evolution of the agent densities is described by the rate equations:
\begin{align}
\begin{split}
\label{eqn:MF}
\dot{L}&=  r  \dot{A} I \equiv (\alpha-1) SI,\\
\dot{I}&= -(1+\gamma r )SI \equiv -\alpha SI ,\\
\dot{S}&= S(I-\gamma),\\
\dot{A}&= \gamma S,
\end{split}
\end{align}
where the dot denotes the time derivative and we define
$\alpha\equiv 1+\gamma r$.  Since the total density is conserved, i.e.,
$L+I+S+A=1$, the sum of these rate equations equals zero.  A natural initial
condition is a population that consists of a small density of susceptible
agents that initiate the dynamics, while all other agents are ignorant; that
is, $I(0)=1-S(0)=I_0$ and $L(0)=A(0)=0$.

To solve these rate equations, it is useful to introduce the modified time
variable $d\tau= S(t)\,dt$, which linearize the rate equations to
\begin{align}
\begin{split}
\label{eqn:RE}
L'&= (\alpha-1) I,\\
I'&=-\alpha I,\\
S'&= I-\gamma,\\
A'&= \gamma,
\end{split}
\end{align}
with solution
\begin{align}
\begin{split}
\label{eqn:sol_tau}
L&= \frac{\alpha-1}{\alpha}I_0(1-e^{-\alpha\tau}),\\
I&= I_0 e^{-\alpha\tau},\\
S&=\frac{I_0}{\alpha}(1-e^{-\alpha\tau})+1-I_0-\gamma\tau,\\
A&= \gamma  \tau.
\end{split}
\end{align}

\begin{figure}[ht]
\centering
\subfigure[]{\includegraphics[width=0.9\linewidth]{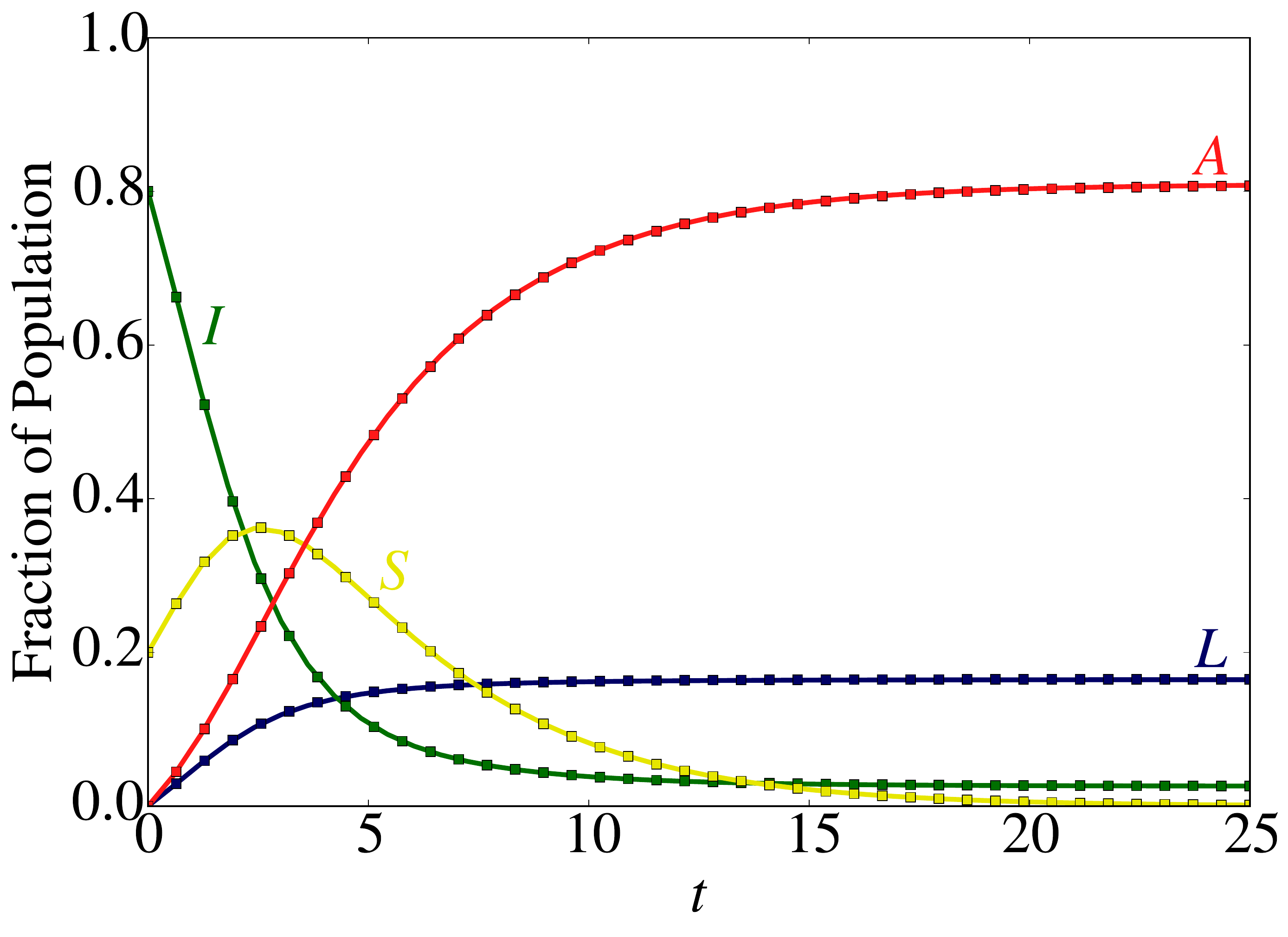}}
\subfigure[]{\includegraphics[width=0.9\linewidth]{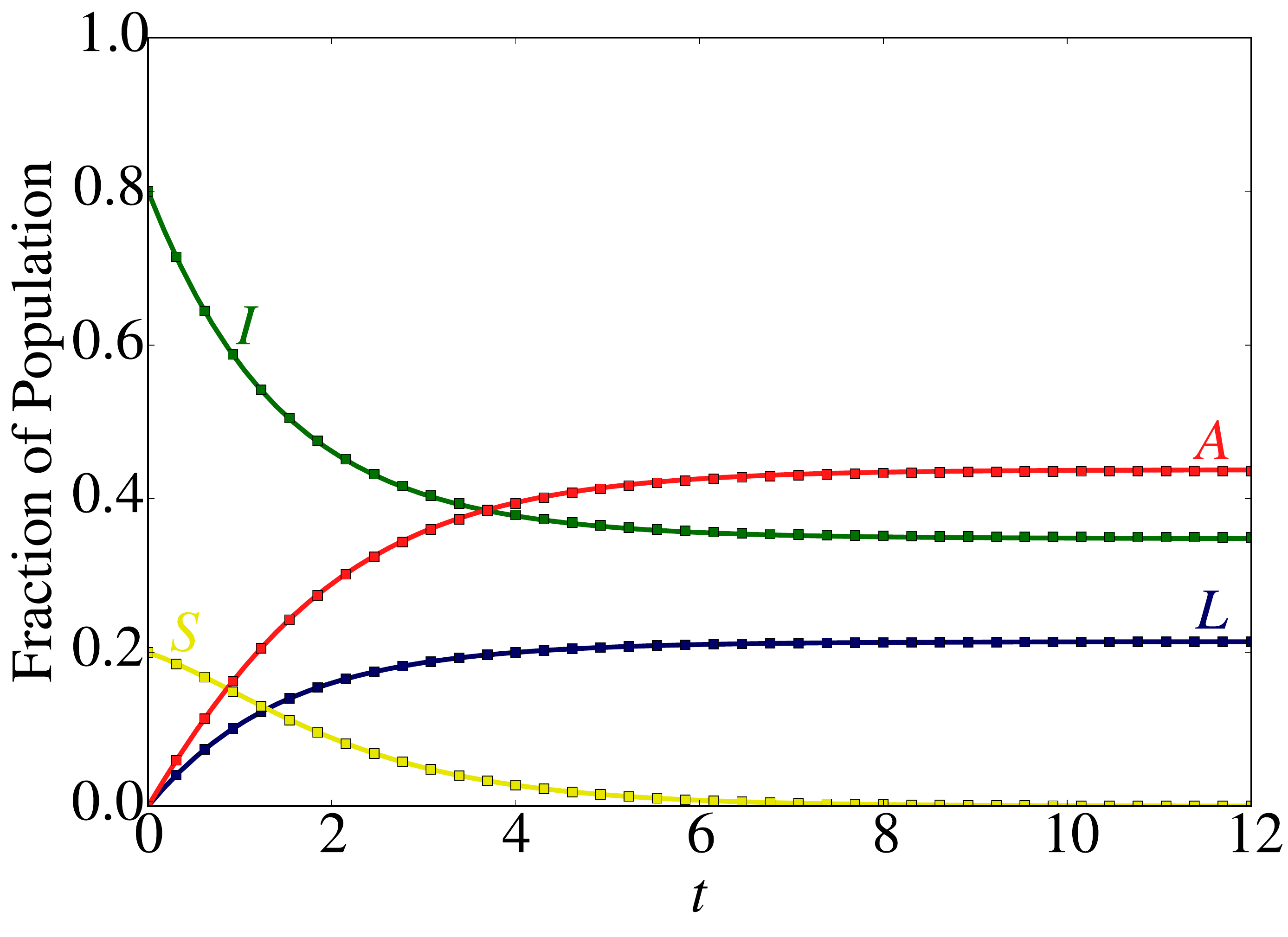}}
\caption{(\textit{Color online}) Evolution of a realization of the LISA model
  on a complete graph of $10^6$ nodes with $I_0=0.8$ and Luddism parameter
  $r=0.9$.  (a) $\gamma = 0.3$ (extensive adoption) (b) $\gamma=1$
  (sparse adoption). Evenly distributed samples of the stochastic simulation ($\Box$) are
  indistinguishable from the solution of Eq.~\eqref{eqn:MF} (solid line).  The
  completion times for (a) and (b) are $60$ and $17$ respectively.}
\label{fig:single}
\end{figure}

There are two basic regimes of behavior that are controlled by the adoption
rate $\gamma$, as illustrated in Fig.~\ref{fig:single}:
\begin{enumerate}
\item[(a)] \emph{Extensive adoption}. When $\gamma<I_0$, the
  density of susceptibles $S$ varies non-monotonically in time and reaches a
  maximum value $S_{\rm inc}$ at an ``inception'' time $t_{\rm inc}$, after
  which $S$ decays to 0.  This non-monotonicity leads to a sigmoidal curve
  for the adopter density, with $A$ increasing rapidly for
  $t\alt t_{\rm inc}$ and increasing very slowly for $t\agt t_{\rm inc}$.
  The rescaled inception time $\tau_{\rm inc}$ is determined by the criterion
  $S'=0$, or equivalently, $I(\tau_{\rm inc})=\gamma$.  This gives
\begin{equation}
\tau_{\rm inc}=\frac{1}{\alpha}\ln(I_0/\gamma)\,.
\end{equation}

\item[(b)] \emph{Sparse adoption}. When $\gamma>I_0$, the
  susceptibles quickly become adopters, leaving behind a substantial static
  population of ignorants and a small fraction of adopters, as well as
  Luddites.
\end{enumerate}
Numerical simulations of the LISA model on a large complete graph \emph{and}
numerical integration of the rate equations \eqref{eqn:MF}, illustrated in
Fig.~\ref{fig:single}, give results that are virtually indistinguishable.

We can express the densities in terms of the physical time $t$ by inverting
$d\tau=S(t)dt$ to give $t=\int_0^{\tau}d\tau'/S(\tau')$.  Substituting
$S(\tau)$ from the third of Eqs.~\eqref{eqn:sol_tau} and taking the limits of
low adoption, $\gamma \ll 1$ and $\alpha\approx1$,
we have~\footnote{Here the term $-\gamma \tau'$ has been neglected.
  This approximation is legitimate since $\tau'$ is integrated from
  $0$ to $\tau\ll \tau_{\infty} \approx 1/\gamma$ and therefore
  $\gamma \tau' \ll 1$ in the regime being considered.  A similar
  reasoning, with $\gamma \tau \leq \gamma \ln(I_0/\gamma) \ll 1$,
  leads to (\ref{eqn:t*_approx}) when $\gamma \ll 1$.}
\begin{equation}
t=\int_0^{\tau}\!\!\frac{ d\tau'}{1-I_0 e^{-\tau'}}\approx \tau +
\ln\left[1-I_0 e^{-\tau}\right]\,.
\end{equation}
In particular, 
the physical inception time $t_{\rm inc}$ is,
\begin{eqnarray}
\label{eqn:t*_approx}
t_{\rm inc}\approx \int_0^{\ln(I_0/\gamma)}\!\!\frac{ d\tau'}{1-I_0 e^{-\tau'}}
\approx \ln\left[\frac{I_0}{(1-I_0)\gamma}\right]
\end{eqnarray}
and therefore grows as $\ln(1/\gamma)$.

The stationary state is reached when all susceptibles disappear, so that no
further reactions can occur.  This gives the criterion $S(\tau_{\infty})=0$
which defines the value of $\tau_{\infty}$.  By solving the third line of
Eq.~\eqref{eqn:sol_tau}, we obtain
\begin{equation}
\label{eqn:Xinf}
\tau_{\infty}=\frac{1}{\gamma}\!-\!\frac{I_0r}{\alpha}\!+\!\frac{1}{\alpha} 
W_0\left(-\frac{I_0}{\gamma} e^{I_0r-\alpha/\gamma}\right)\,,
\end{equation}
where $W_0(z)$ is the principal branch of the Lambert function $W(z)$, which
is defined as the solution of $z=We^{W}$.  Here $\tau_{\infty}$ is a
decreasing function of the adoption rate $\gamma$, with $\tau_{\infty} \sim 1/ \gamma$ in the high and low adoption rate regimes.

We now determine the final densities by substituting $\tau_\infty$ into
Eqs.~\eqref{eqn:sol_tau}.  For small adoption rate ($\gamma \ll 1$), this
gives 
\begin{align}
A_\infty&=1\!-\!\mathcal{O} (\gamma),\nonumber\\
I_\infty&\to 0,\nonumber\\
L_\infty &\approx (\alpha-1) I_0=\mathcal{O} (\gamma).\nonumber
\end{align}  
Similarly, the densities at the inception time are obtained by substituting $\tau_{\rm inc}$
into Eqs.~\eqref{eqn:sol_tau}. This yields $A(\tau_{\rm inc}) + S(\tau_{\rm inc}) =
1-[(\alpha-1)I_0+\gamma]/\alpha$. Since $(\alpha-1)I_0 \sim {\cal O}(\gamma)$, when $\gamma \ll 1$ and $r$ is finite, here the stationary density of adopters  approximately equals the sum of the adopter and susceptible densities at the inception time, $A_\infty \approx A(\tau_{\rm inc}) + S(\tau_{\rm inc})$. Hence, in the low adoption rate regime (when $r$ is finite), we can infer the final level of adoption from the adopter and susceptible densities at the inception time, i.e., well before the stationary state.
 
\begin{figure}[ht]
\centering
\includegraphics[width=0.8\linewidth]{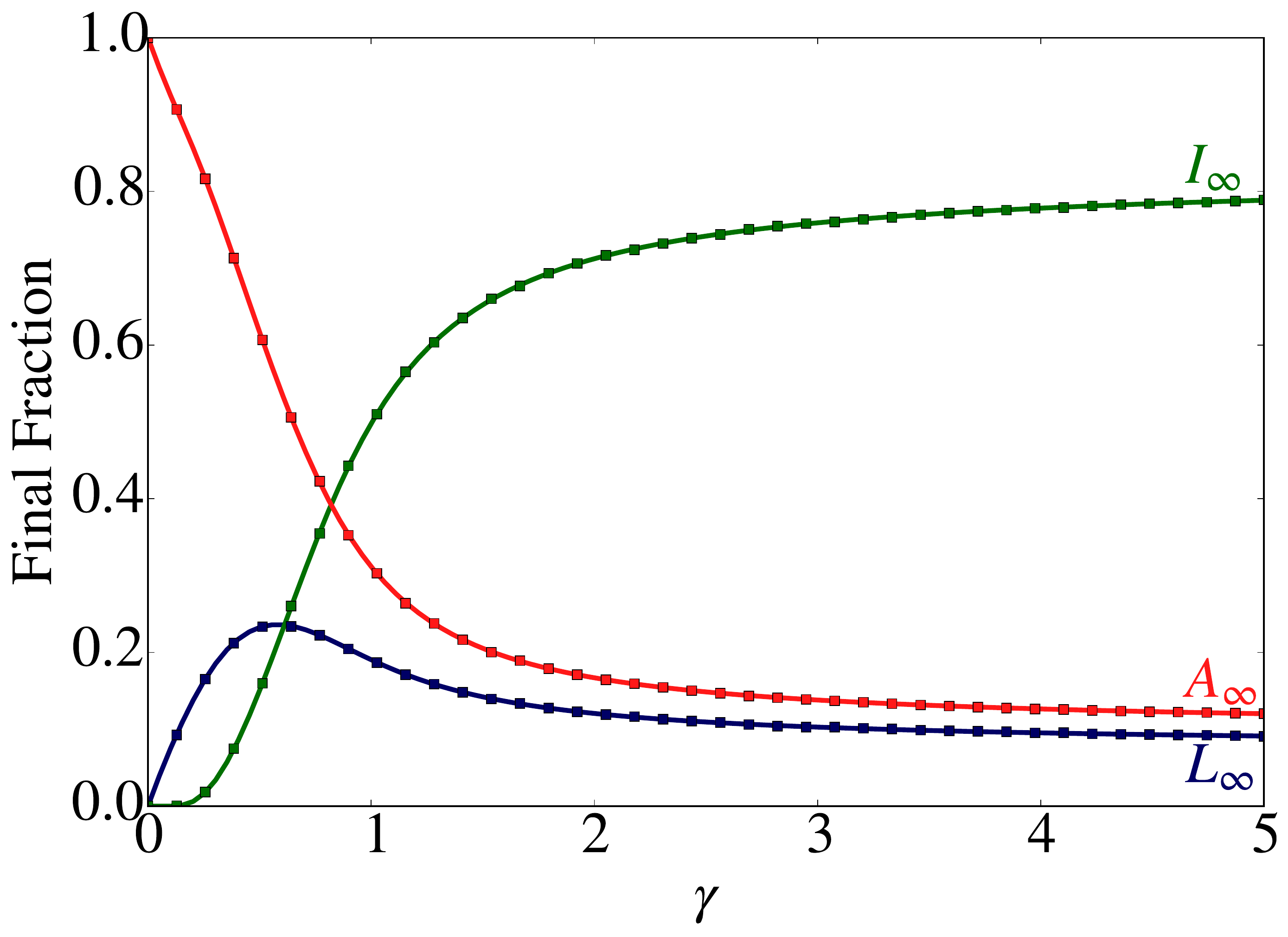}
\includegraphics[width=0.8\linewidth]{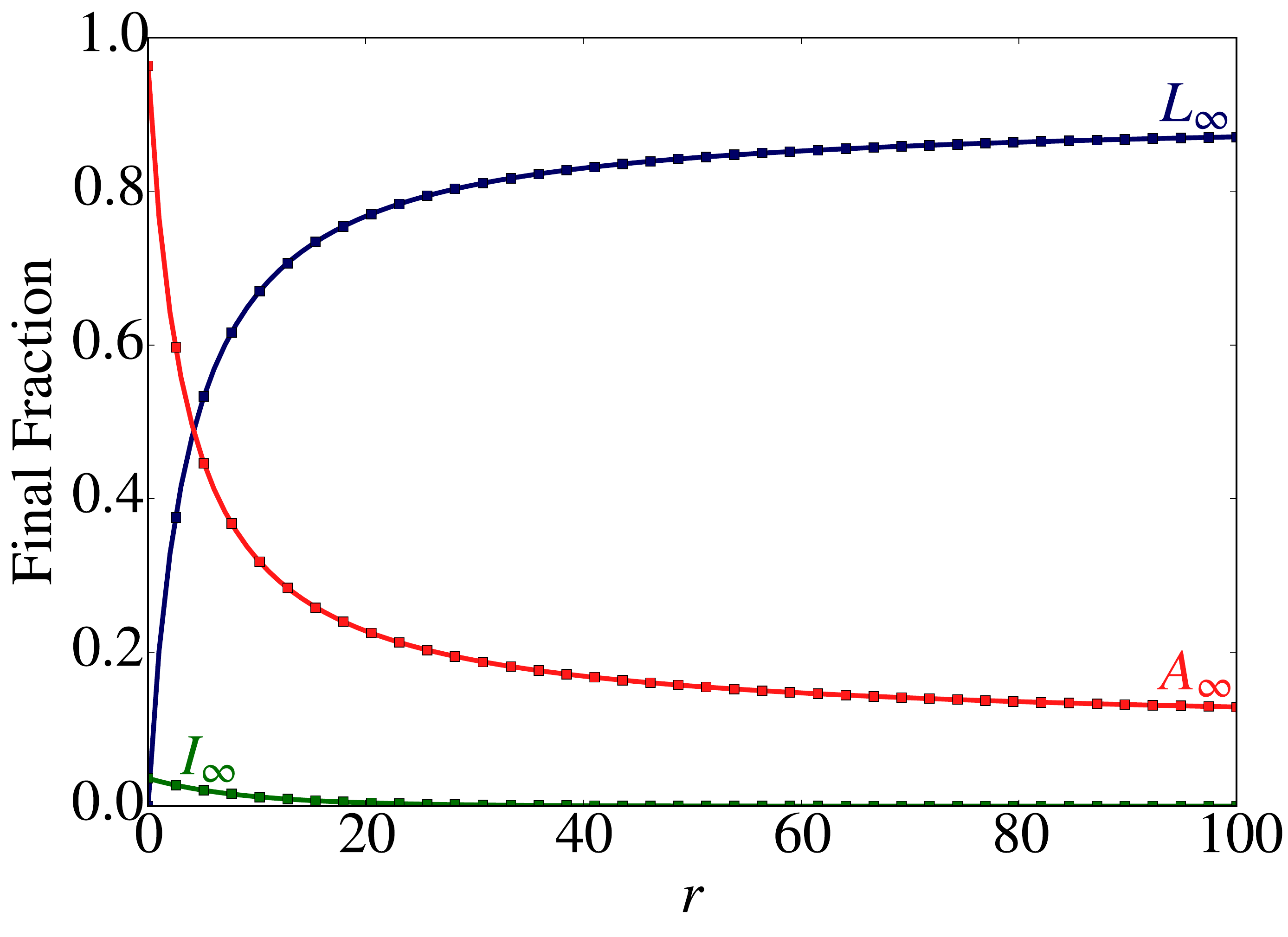}
\caption{(\textit{Color online}) Dependences of the final-state densities
  $L_\infty, I_\infty$ and $A_\infty$ for a complete graph of $10^4$ nodes
  and $I_0=0.9$. In the top panel $r=0.9$ while $\gamma$ varies, whereas in
  the bottom panel $\gamma=0.3$ while $r$ varies. Simulations ($\Box$) in complete agreement with \eqref{eqn:sol_tau} with substitution \eqref{eqn:Xinf} (solid line). }
\label{fig:parameters}
\end{figure}

The dependence of
the final densities for different parameter ranges is shown in
Fig.~\ref{fig:parameters}.  Again simulation results for the complete graph
are indistinguishable from numerical integration of the rate equations.
Interestingly, $L_\infty$ varies non-monotonically on $\gamma$ when the
initial state consists mostly of ignorants and the fixed rate of Luddism $r$
is not too high, as in Fig.~\ref{fig:parameters}~(top).
This non-monotonic dependence on $\gamma$ can be understood by noting 
that $dL_\infty/d\gamma \sim r(1 - e^{-1/\gamma}) > 0$ for $\gamma \ll 1$ and $dL_\infty/d\gamma \sim -e^{-1/\gamma}/\gamma^2 < 0$
for $\gamma \gg 1$.
We therefore expect that $L_\infty$ is peaked for an intermediate value of $\gamma$  on a range between the slow  and quick adoption regimes.  It is also worth
noting that in the absence of Luddites, complete adoption is almost, but not
completely achieved, since the final densities of adopters and ignorants are
$A_\infty\approx 1- I_\infty$ and $I_\infty \approx e^{-1/\gamma}$, see Fig.3 (bottom).

To assess the role of finite-$N$ fluctuations on the dynamics, we simulate
the LISA model on complete graphs of $N$ nodes using the Gillespie
algorithm~\cite{Gillespie}.  At long times we find that the densities of each
species, $N_{X}/N$, fluctuates around the corresponding mean-field density,
with a root-mean-square fluctuation of amplitude $\sim N^{-1/2}$, as
expected from general properties of this class of reaction
processes~\cite{noise}.  We also find that the probability distribution of
$N_{X}/N$ is a Gaussian of width of order $N^{-1/2}$ that is centered on the
mean-field density.  We also estimate the completion time $T_C$ for the
system to reach its final state by the physical criterion that
$S(t\!=\!T_C)=1/N$.  That is, completion is defined by the presence of a
single susceptible remaining in the population~\cite{reinf}.  By linearizing
the rate equations \eqref{eqn:MF} around $S_\infty=0$, the density of
susceptibles asymptotically vanishes as $S(t)\sim e^{-(\gamma-I_\infty)t}$.
Hence, we estimate the mean completion time to be
$T_C \approx (\ln~N)/(\gamma-I_\infty)$.  This prediction is confirmed by our
simulations.

\section{LISA model on random graphs and lattices}
\label{sec:graphs}

We now consider the behavior of the LISA model on Erd\H os-R\'enyi random
graphs and one-dimensional lattices.  We are particularly interested in
uncovering dynamics that are characterized by genuine non mean-field effects.

A graph with $N$ nodes can be represented by its $N\times N$ adjacency matrix
${\bf A}=[A_{ij}]$, where $A_{ij}=1$ if nodes $i$ and $j$ are connected and
$0$ otherwise.  We implement the LISA model on such a graph using the
Gillespie algorithm~\cite{Gillespie}.  The propensity for a susceptible to
become an adopter is $\gamma$, independent of the local environment.  The
propensity for an ignorant node $i$ to become susceptible if it has $s_i$
susceptible neighbors is $s_i/N$.  The propensity of an ignorant node $i$
to become a Luddite is $r\gamma s_i/k_i$, where $k_i=\sum_j A_{ij}$ is the
degree (number of neighbors) of node $i$, and $s_i/k_i$ is the fraction of
nodes in the neighborhood of $i$ that are in the susceptible state.  Thus the
propensity of $i$ to become a Luddite is proportional to the sum of its
susceptible neighbors' propensities to adopt.  This rate encodes node $i$'s
local knowledge of the rate of adoption.  These reaction rates approach those
of the complete graph, described in Section~\ref{sec:MF}, as the average
degree of the graph increases.

\subsection{Erd\H os-R\'enyi random graphs}
\label{sec:ER}
We first study the LISA model on the class of Erd\H os-R\'enyi (ER) random
graphs in which an edge between any two nodes occurs with a fixed probability
$p$.  This construction leads to a binomial degree distribution for the ER
graph in which each node has, on average, $k = p(N-1)$
neighbors~\cite{Newman2010}.  Under the assumption of no correlations between
the degrees of neighboring nodes, the adjacency matrix may be written as
$A_{ij}\approx k_i k_j /(N k) \approx k/N$.  The LISA dynamics on ER graphs
can now be approximately described by a natural generalization of the
mean-field theory in which there are suitably defined reaction rates.  In
particular,
if $S_i$ is the probability that a node $i$ is susceptible and $I_j$
is the probability that a node $j$ is ignorant, then the density of
susceptibles $S$ evolves as
\begin{align*}
\dot{S_i}=S_i\Big[\sum_j (A_{ij}/N)I_j -\gamma \Big]\approx S\big[(k/N)I-\gamma\big],
\end{align*}
since each susceptible interacts with $k$ of its $N$ neighbors on average.
Thus on the ER graph there is a rescaling of the rate of the two-body
contagion process $\mathcal{I}+ \mathcal{S} \to \mathcal{S}+\mathcal{S}$,
whereas the rates of the remaining one-body processes remain unaltered.
Hence we obtain the effective rate equations
\begin{align}
\begin{split}
\label{eqn:MF_ERG}
\dot{L}&= \gamma r  S I \equiv \left(\beta - \frac{k}{N}\right) SI,\\
\dot{I}&= -\left(\gamma r +\frac{k}{N}\right)S I \equiv -\beta SI,\\
\dot{S}&= S\left(\frac{k}{N}~I-\gamma\right),\\
\dot{A}&= \gamma S,\\
\end{split}
\end{align}
where, for later  convenience, we define $\beta\equiv \gamma r + (k/N)$.

As in the case of the mean-field dynamics, the above equations predict two
regimes of behavior (see Fig.~\ref{fig:er_single}):
\begin{figure}[ht]
\centering
\subfigure[]{\includegraphics[width=0.8\linewidth]{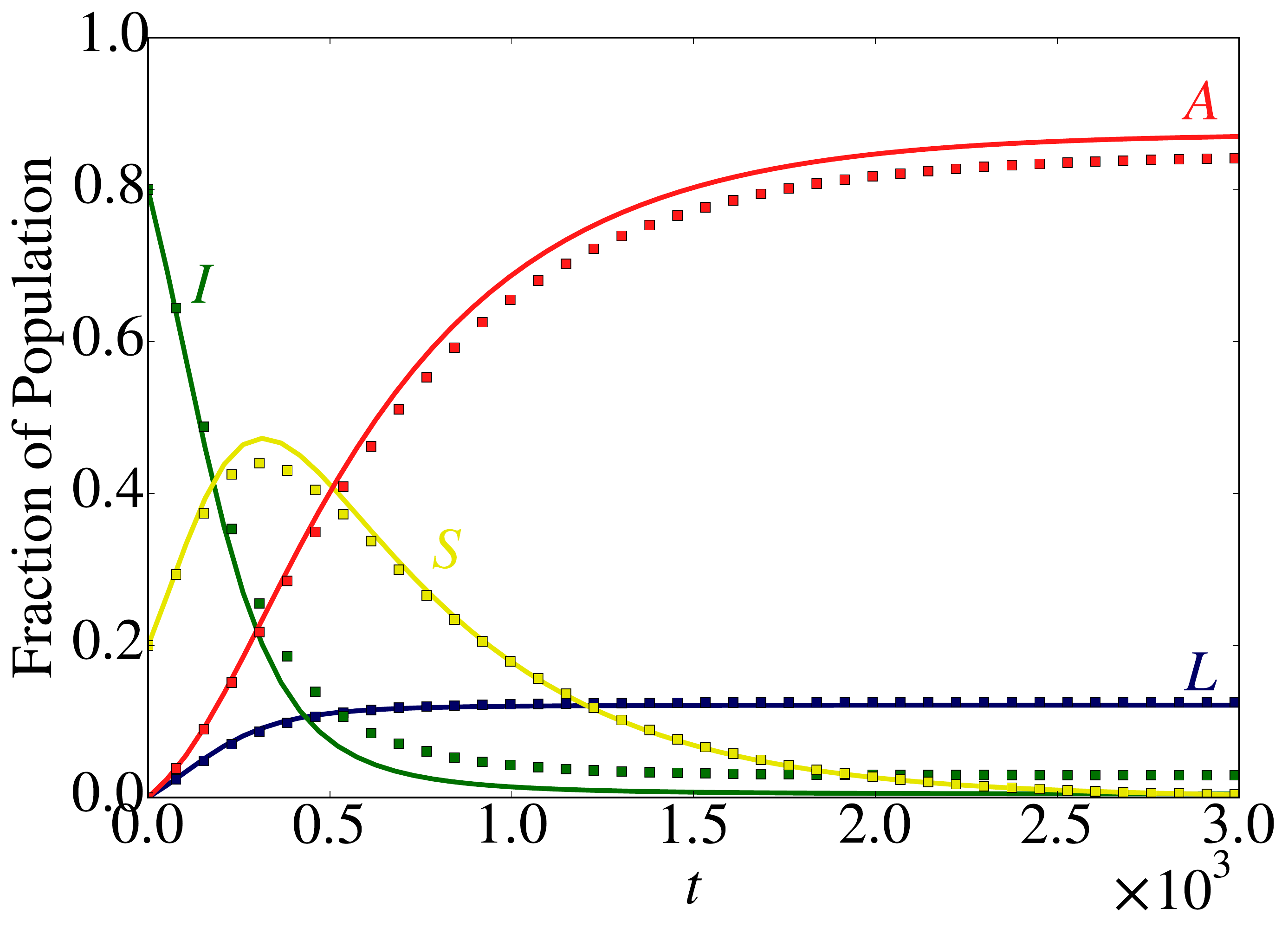}}
\subfigure[]{\includegraphics[width=0.8\linewidth]{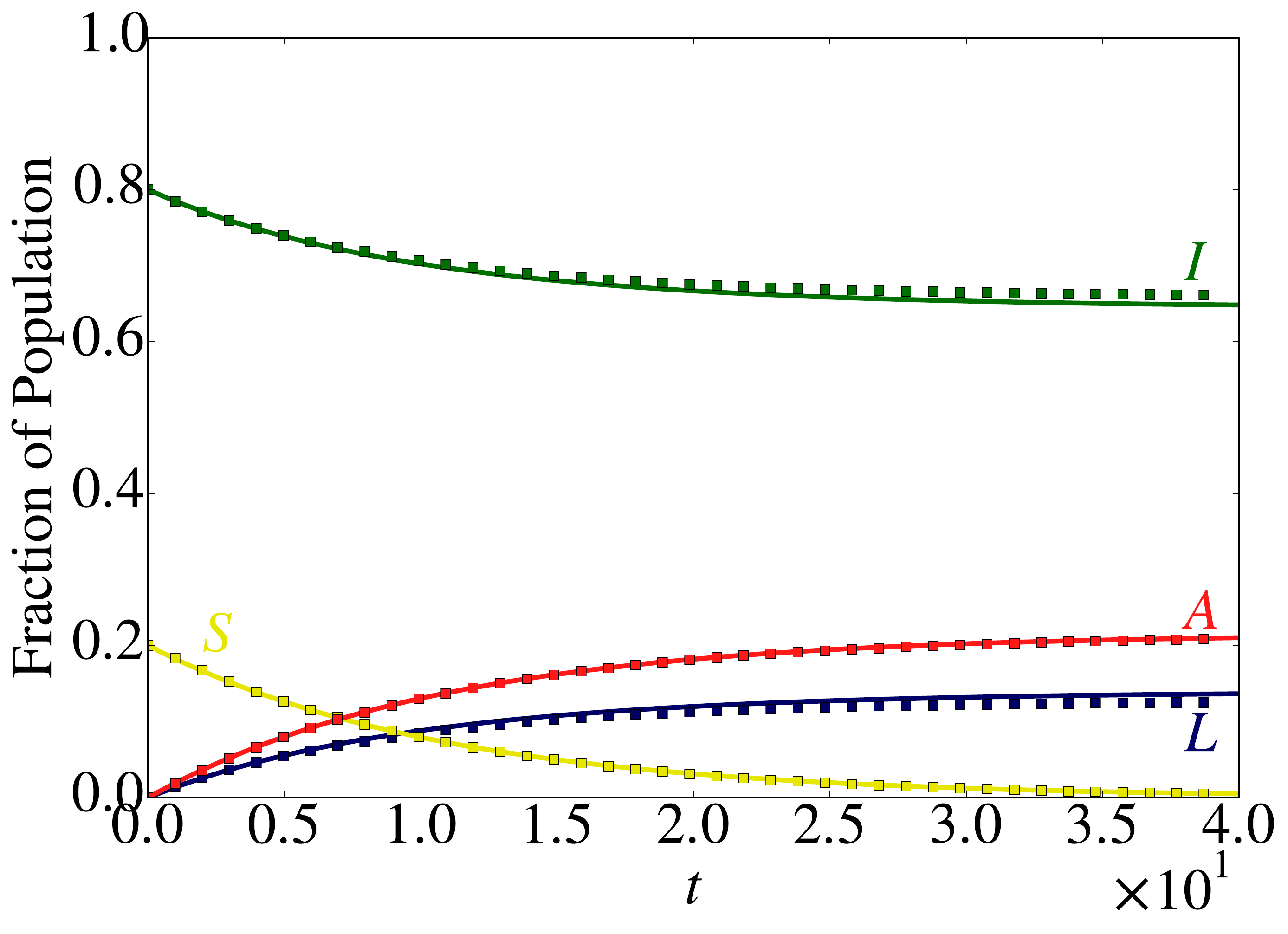}}
\caption{(\textit{Color online}) The evolution, averaged over 100
  realizations, of the LISA model on an ER graph with $N=10^3$ nodes, $k=10$, and  $I_0=0.8$. (a) $\gamma = 0.002$, such that $\gamma < (k/N) I_0$
  and (b) $\gamma = 0.1$ such that $\gamma > (k/N) I_0$.  Shown are the
  evenly distributed samples of the stochastic simulation ($\Box$) and the solution of Eq.~\eqref{eqn:MF_ERG} (solid line).
  The Luddism parameter $r=0.9$.}
\label{fig:er_single}
\end{figure}
\begin{enumerate}
\item[(a)] {\it Slow but extensive adoption} ($\gamma<kI_0/N$).  Here the
  density of $\mathcal{S}$'s peaks at a inception time
  $t_{\rm inc}\sim \ln(1/\gamma)$ before vanishing.

\item[(b)] {\it Rapid but sparse adoption} ($\gamma>kI_0/N$).  The density of
  $\mathcal{S}$'s vanishes quickly so that the density of adopters and
  Luddites quickly reach their steady-state values.
\end{enumerate} 

The simulation results presented in Fig.~\ref{fig:er_single} indicate
that the mean-field approximation \eqref{eqn:MF_ERG} correctly
captures the main qualitative features of the dynamics on large ER
graphs. 
When $\gamma <kI_0/N$, the densities of $\mathcal{A}$ and $\mathcal{L}$ are
characterized by a sigmoidal time dependence, whereas the density of
$\mathcal{S}$ has a peak at the inception time $t_{\rm inc}$, with time
evolution that is slower than on complete graphs, since each agent has now a
finite neighborhood.

The stationary state can be determined by again noting that
\eqref{eqn:MF_ERG} becomes linear in terms of the variable
$\tau=\int_0^t S(t') dt'$.  Thus proceeding as in Section~\ref{sec:MF}, we find the
steady state by the condition $S(\tau_{\infty})=0$. This yields
\begin{align}
\begin{split}
\label{eqn:sol_tau_ERG}
I_\infty&= I_0 e^{-\beta\tau_{\infty}}\\
L_\infty&= \frac{\beta - k/N}{\beta} (I_0-I_\infty)\\
A_\infty&= \gamma \tau_{\infty},\\
\end{split}
\end{align}
where now
\begin{align}
\label{eqn:MF_Xinf}
\tau_{\infty}&=\frac{k}{N\gamma\beta}+(1-I_0)\frac{r}{\beta}  \nonumber \\
&\hskip 0.5in + \frac{1}{\beta} W_0\left(-\frac{kI_0}{N\gamma}e^{-(1-I_0)r-k/(N\gamma)}\right)\,.
\end{align}
\begin{figure}[th]
\centering
\includegraphics[width=0.8\linewidth]{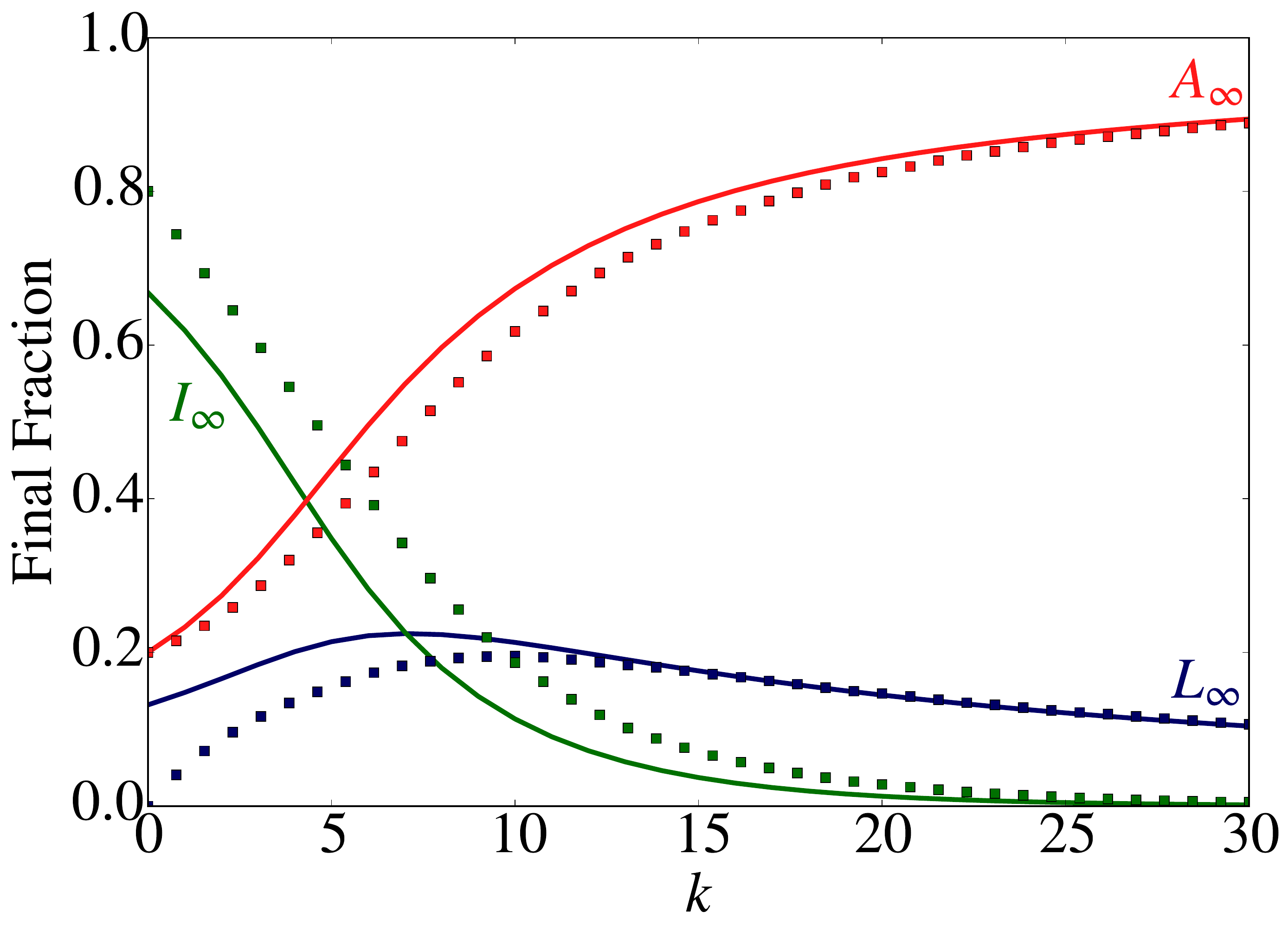}
\caption{(\textit{Color online}) Dependence of the final densities
  $L_\infty, I_\infty$ and $A_\infty$ on the average degree for ER graphs
  with $N=10^3$ nodes.  The simulation ($\Box$) represents an average over
  $40$ model realizations for $30$ randomly generated networks. Parameters
  are $\gamma = 0.005, r=0.9$, and $I_0=0.9$. The mean-field predictions  \eqref{eqn:sol_tau_ERG} (solid line) match the simulation for $k \gtrsim 20$ (see main text).}
\label{fig:er_parameters}
\end{figure}

It is instructive to compare the predictions \eqref{eqn:sol_tau_ERG} with the
results of stochastic simulations, and also compare with the equivalent
results for the complete graph.  Figure~\ref{fig:er_parameters} shows
simulation results for the stationary densities as a function of the mean
degree.  These results confirm that the
mean-field predictions correctly capture the functional dependence of the
steady state on the various parameters. However, the
mean-field predictions \eqref{eqn:sol_tau_ERG} are quantitatively accurate
only when $k/N$ is large.  If $k/N\ll 1$, the neighborhood of each agent
represents a small fraction of the entire network, and resulting large
demographic fluctuations invalidate the assumptions underlying the derivation
of \eqref{eqn:sol_tau_ERG}. The dependence on $\gamma$ and $r$
are qualitatively similar to those observed on complete graphs.

The influence of demographic fluctuations can be heuristically assessed by
viewing ER graphs of mean degree $k$ as a meta-population that consists of
$N/k$ patches each comprising a well-mixed population of size $k$.  According
to this picture, when $N\gg k\gg 1$, the number of agents in each of the
$N/k$ components fluctuates in a range of $k^{1/2}$ about its average value.
Since these are independent fluctuations, the noise in the whole population
should have an amplitude $\sim (N/k)^{1/2}\,k^{1/2}=N^{1/2}$, which leads to
fluctuations in the densities of order $N^{-1/2}$.  This prediction is
confirmed by our simulations---we find that $N_A(\infty)/N$ has a Gaussian
probability distribution around $A_\infty$ with a width that decays as
$N^{-1/2}$. The same behavior is observed for $L_\infty$ and $I_\infty$ but not for $S_\infty$ as $S_\infty=0$ is a requirement for the completion of the dynamics.

\begin{figure}[ht]
\includegraphics[width=0.9\linewidth]{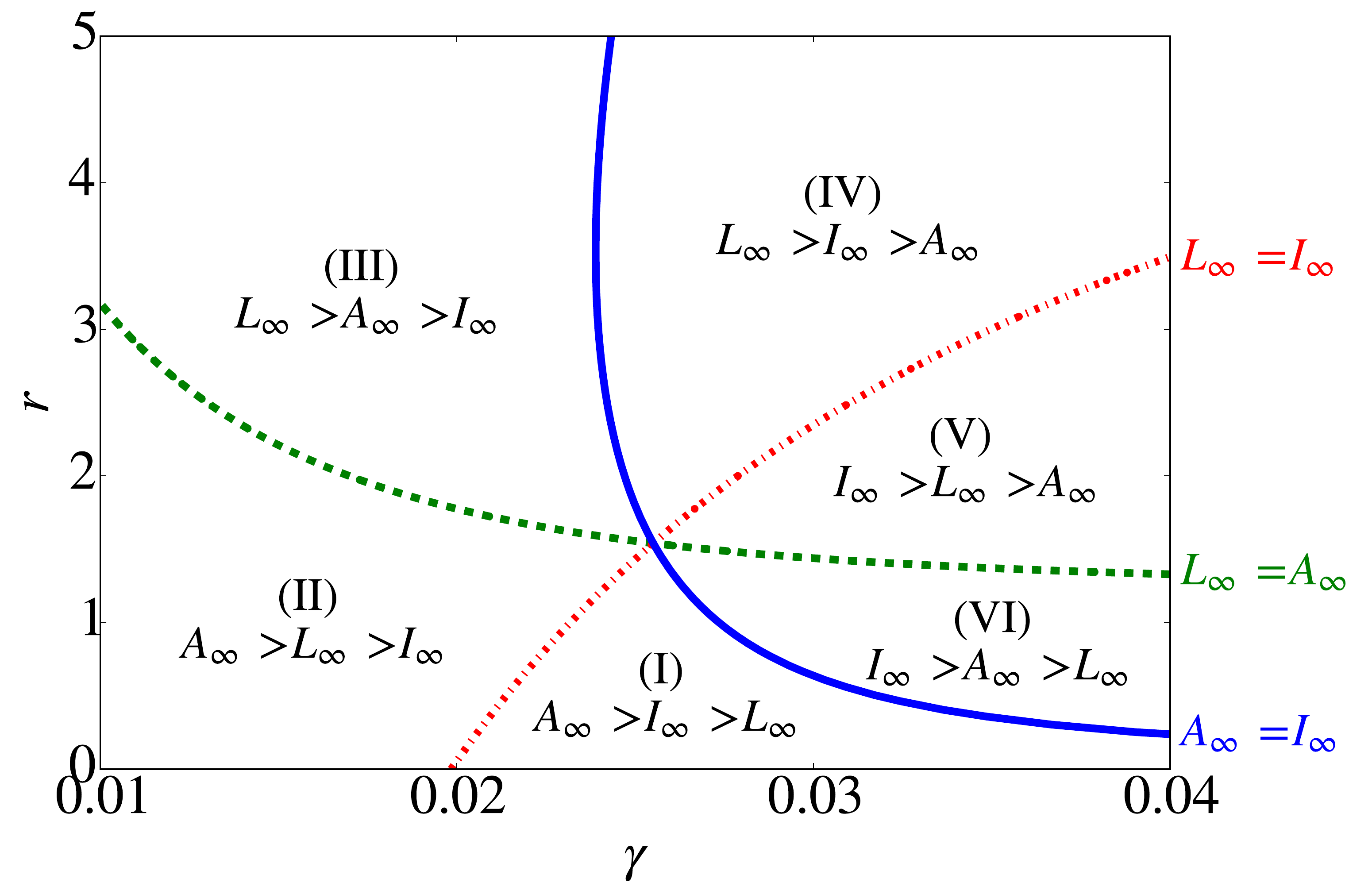}
\caption{(\textit{Color online}) The mean-field steady state predictions \eqref{eqn:sol_tau_ERG} over the parameter space $(\gamma,r)$ for $k/N = 0.025$ and $I_0=0.9$. The contours $L_\infty = I_\infty$, $L_\infty = A_\infty$, and $I_\infty=A_\infty$ split the domain into six regions which characterize the innovation (see main text).}
\label{fig:phase}
\end{figure}

The mean-field steady state predictions \eqref{eqn:sol_tau_ERG} are 
summarized in Fig.~\ref{fig:phase}, where we plot the mean-field steady state predictions
corresponding to each pair of steady state densities being equal. For 
example, the solid curve corresponds to parameter values for which 
$A_\infty=I_\infty$. This carves up the $(\gamma,r)$ parameter space 
into regions corresponding to different orderings of the steady state 
densities, labelled with Roman numerals in Fig.~\ref{fig:phase}. We 
can use these orderings to interpret, from a marketing perspective, 
whether or not these would be considered successful campaigns. In this 
context, the most desirable outcome would be region (I), where 
adopters form the largest steady-state group and Luddites form the 
smallest. Whilst adopters also form the largest group in region (II), 
Luddites form the second largest group and so this region could be 
considered a controversial success --- despite the majority adopting, 
a significant number of people have responded negatively. Conversely, 
regions (III) and (IV) could both be considered controversial failures 
because Luddites form the largest groups. Regions (V) and (VI) 
represent ineffective campaigns because ignorants form the largest 
steady state groups.

In summary, we have shown that the LISA dynamics on ER graphs can be
accurately approximated by using mean-field assumptions, provided that the
average degree is sufficiently high (see Fig.~\ref{fig:er_parameters}).

\subsection{One-dimensional lattices}
\label{sec:1D}
Recent controlled experiments have shown that innovation may spread
more efficiently on clustered graphs and lattices than on random
networks~\cite{C10}.  To understand the effect of regular topology
on the spread of an innovation, and where the mean-field approximation
breaks down, we investigate the LISA dynamics on one-dimensional
lattices.

The two regimes of behavior predicted by the mean-field description
\eqref{eqn:MF_ERG} on ER random graphs (see Section~\ref{sec:ER}) also occur
on one-dimensional lattices, despite the difference in
topology. Specifically with $k=2$ we observe slow adoption for
$\gamma<(2/N)I_0$ and fast adoption for $\gamma>(2/N)I_0$. 
From simulations, illustrated in Fig.7, we observe the following three regimes:

\begin{enumerate}
\item[(A)] When $\gamma\ll 2I_0/N$, there is slow adoption as well as a
  time-scale separation.  First, almost all $\mathcal{I}$'s are converted to
  $\mathcal{S}$'s~\cite{1D} in a time of the order of $N^2$.  When the
  lattice consists almost entirely of $\mathcal{S}$'s, these become adopters
  after a mean time of the order of $\gamma^{-1}$.  As a consequence, when
  $\gamma\ll N^{-1}$ the size of the adopter domains grows abruptly after a
  time of order $\sim N^2 +\gamma^{-1}$, when all ignorants have disappeared
  and the entire lattice is covered with adopters.

\item[(B)] When $\gamma \sim 2I_0/N$, the domains of adopters grow initially
  nearly linearly in time, whereas the average size of ${\cal I}$ clusters
  remains approximately constant and of a comparable size to ${\cal A}$
  domains.

\item[(C)] When $\gamma\gg (2/N)I_0$, adoption occurs quickly and the final
  state is reached in a time of order ${\cal O}(1/\gamma)$.  The final
  adopter density is limited by the formation of Luddites at the ends of ignorant domains which prevent further conversion within each domain.
\end{enumerate}
\begin{figure}[ht]
\includegraphics[width=0.8\linewidth]{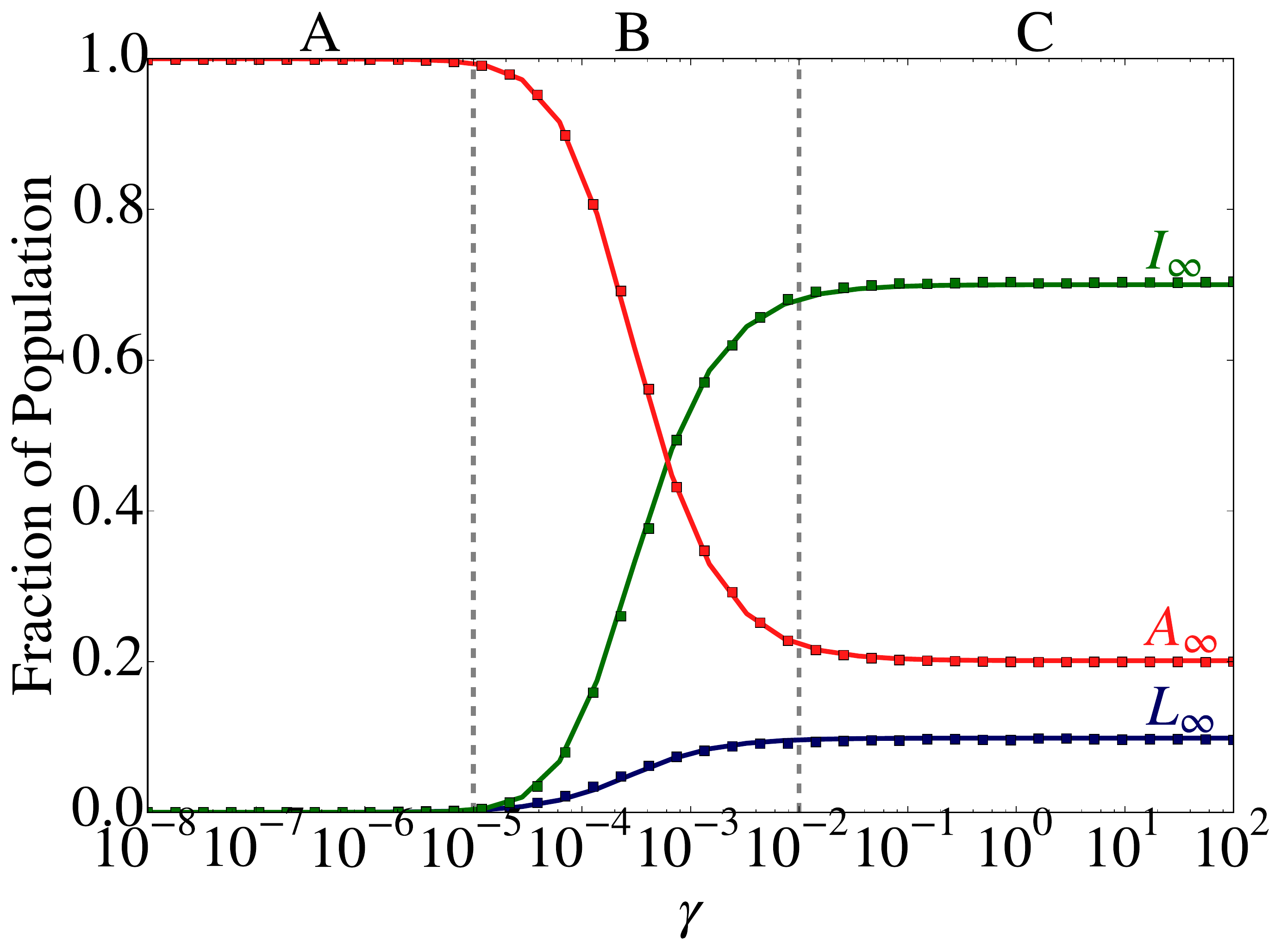}
\includegraphics[width=0.9\linewidth]{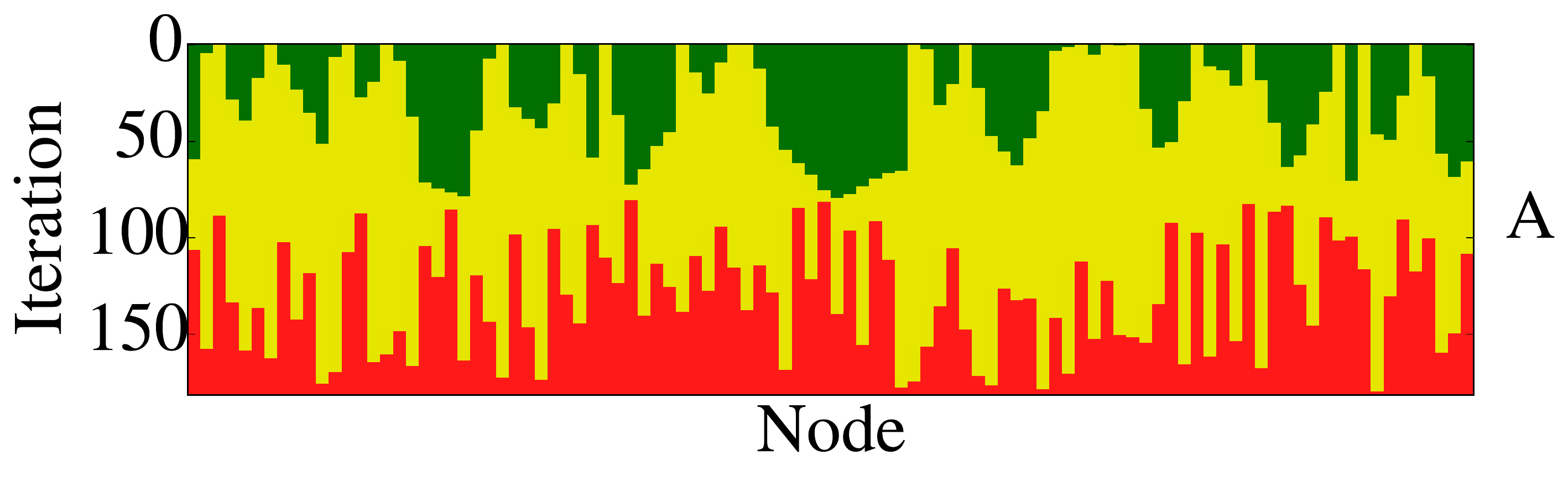}
\includegraphics[width=0.9\linewidth]{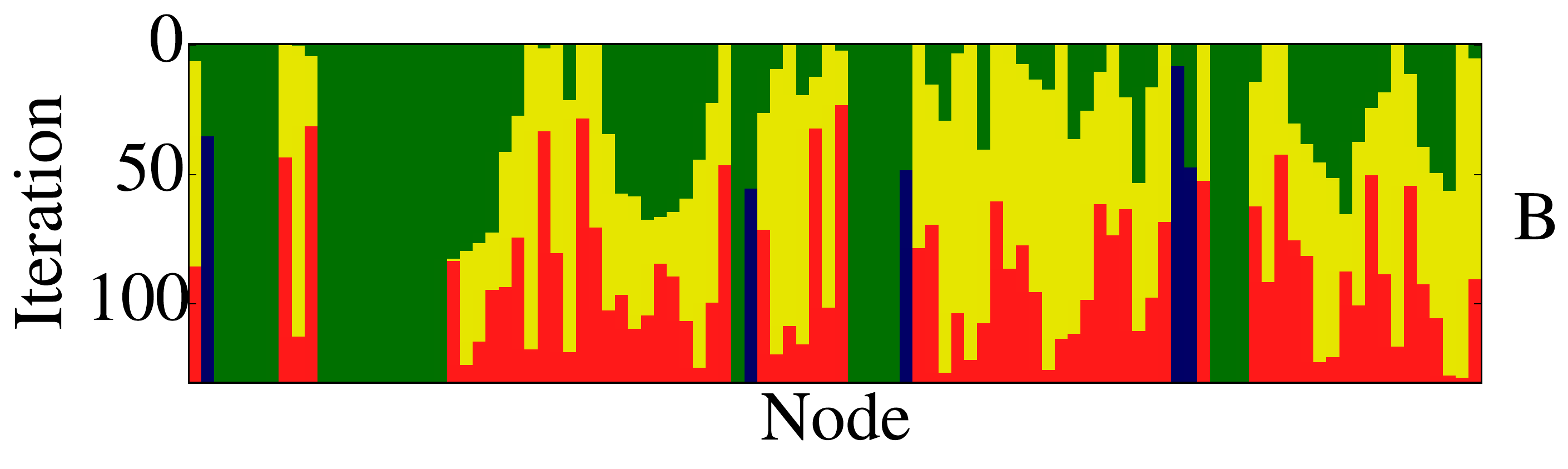}
\includegraphics[width=0.9\linewidth]{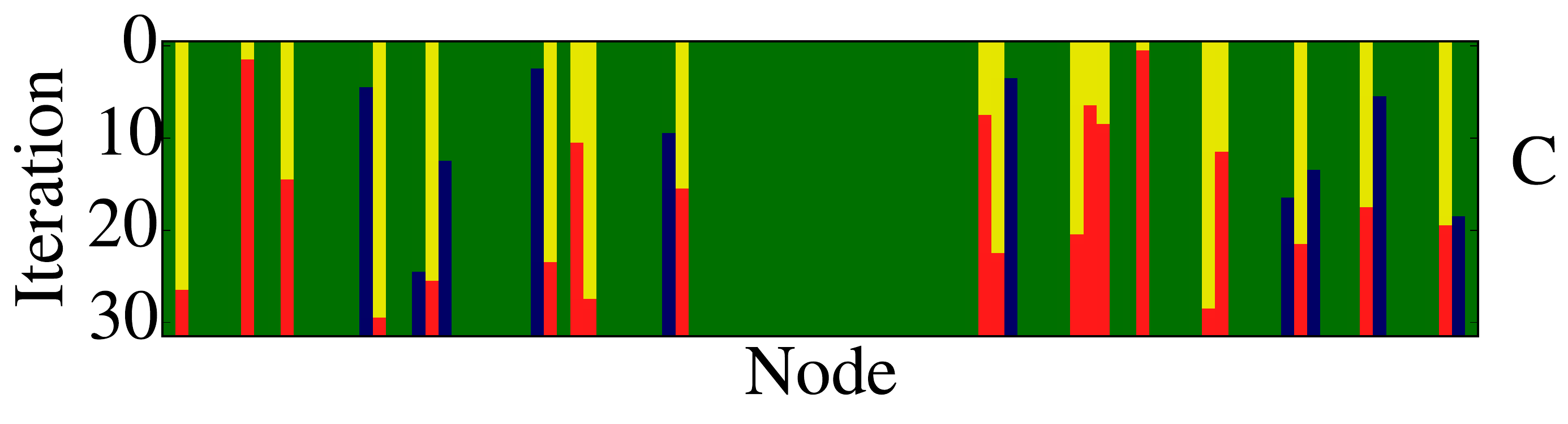}
\caption{(\textit{Color online}) Final simulated average proportions of
  adopters (red/gray $\Box$), ignorants (green/dark gray $\Box$) and Luddites (blue/black $\Box$) for
  varying values of $\gamma$, averaged over 100 simulations. Theoretical
  predictions using ignorant domain length (see Appendix~\ref{ap:1D} for
  details) are overlaid (solid line). Parameters are $N=1000$, $r=0.5$. Initially
  ignorants and susceptibles are randomly distributed, with densities
  $I_0 = 0.8$ and $S_0 = 0.2$. The three regimes discussed in the text are
  separated by dashed lines corresponding to regions where
  $(2/N)I_0\ll \gamma$ and $(2/N)I_0\gg \gamma$. Typical realizations of the
  model for $N=100$ in each of the three regimes are given (bottom). On the
  vertical axis the iteration corresponds to a single step of the Gillespie
  algorithm, with one reaction taking place per iteration.}
\label{fig:1dlattice_gamma}
\end{figure}

\begin{figure}[ht]
\centering
\includegraphics[width=0.9\linewidth]{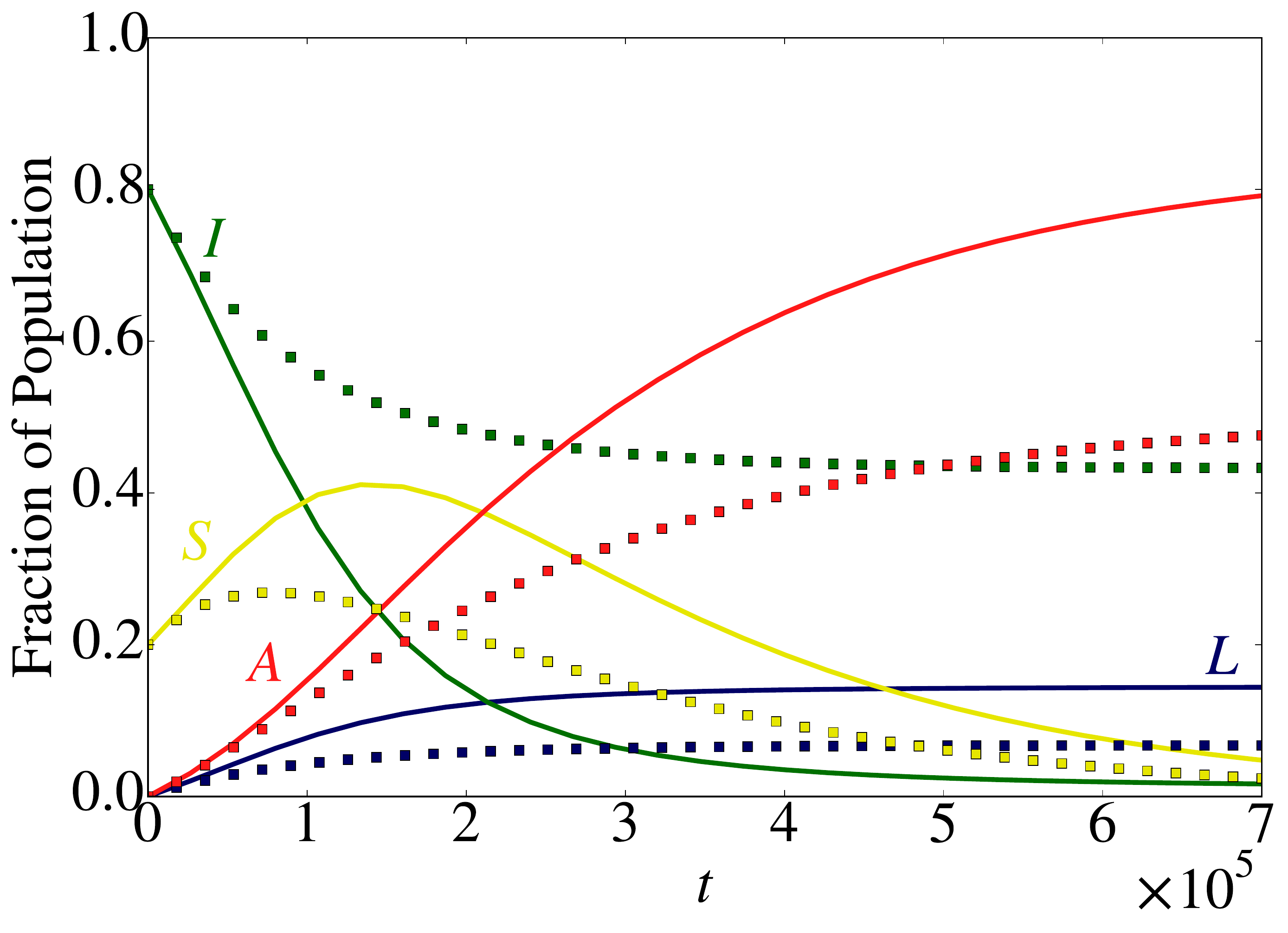}
\caption{(\textit{Color online}) Time dependence of the densities in each
  state for a one-dimensional lattice of size $N=10^5$ averaged over 100
  realizations.  The corresponding mean-field predictions from
  Eq.~\eqref{eqn:MF_ERG} with $k=2$ (solid line) deviate dramatically from the simulation samples ($\Box$).  The parameters are
  $\gamma = 0.005, r = 0.9$, and $I_0 = 0.8$.}
\label{fig:lattice_single}
\end{figure}

While the mean-field approximation \eqref{eqn:MF_ERG} predicts the correct
regimes of behavior, the agreement is only qualitative.  In
Fig.~\ref{fig:lattice_single} we compare typical simulations of the LISA
model on a one-dimensional lattice with the mean-field predictions of
\eqref{eqn:MF_ERG} for the case of $k=2$.
The simulations and mean-field predictions \eqref{eqn:MF_ERG} systematically
deviate; the latter always overestimates the final density of adopters and
underestimates the final density of ignorants. This can be attributed to the
topological constraints on one-dimensional lattices.  Initially the lattice
comprises of contiguous domains of ignorants that are separated by domains of
one or more neighboring susceptibles.  Since ignorants can only become
susceptible if a neighbor is susceptible, domains of ignorants shrink at
their interfaces with susceptibles.  Crucially, the evolution of an
ignorant-susceptible interface ceases if either the susceptible at the
interface adopts or the ignorant at the interface becomes a Luddite.  Thus in
one dimension both Luddites and adopters act as barriers to the spread of
adoption, an effect that is not captured by the mean-field description.

Since domains of ignorants decrease in size and evolve independently, we can
determine analytically the expected final length of ignorant domains
$\langle x\rangle$ and hence the final fractions of each type of agent.  The
details of these calculations are given in Appendix~\ref{ap:1D}. Briefly, we
first determine the probability $P_n(m)$ that a domain of ignorants of
initial length $n$ shrinks by $m$.  We then use $P_n(m)$ to calculate the
expected final length of ignorant domains $\langle x\rangle$ and the final
fraction of ignorants.  Since Luddites only form at the boundaries of
ignorant domains, we are able also to determine the expected final fraction
of Luddites and hence, using the conservation relation $L+I+S+A=1$, the final
fraction of adopters.  The resulting final fractions of each type of agent
are plotted in Fig.~\ref{fig:1dlattice_gamma} and agree extremely well with
the corresponding numerical simulations.  In principle, this method allows us
to derive explicit formulas for the final fractions of each agent; however,
in practice these formulas prove cumbersome.

\section{Discussion \& conclusion}
\label{sec:conc}

Innovations are often accompanied by societal debates and controversies that
may lead to divisions between adopters of an innovation and those who
permanently reject that innovation.  Consequently, innovations are rarely
adopted by the whole population, as various examples, ranging from technology
to medicine, demonstrate.  Classical models of innovation diffusion, such as
that proposed by Bass, assume a ``pro-innovation bias'' and predict the
complete adoption of innovations.

Motivated by these considerations we have introduced a multi-stage
generalization of the Bass model, the LISA model, that does not unavoidably
lead to complete adoption.  The main new feature of our model is the
introduction of Luddites that permanently oppose the spread of innovation in
their neighborhood. In the LISA model, ignorant individuals can successively
become susceptibles and then adopters, or turn to Luddism in response to a
high rate of adoption and permanently reject the innovation.

We carried out a detailed analysis of the properties of the LISA model on
complete graphs and Erd\H os-R\'enyi random graphs, as well as on
one-dimensional lattices. In particular, we focused on the steady states and
completion time (time to reach stationarity).  We showed that significant
insights can be gained from a simple mean-field analysis that aptly captures
the qualitative aspects of the two basic regimes of the LISA dynamics.  When
the rate of adoption is low, the population slowly converges to a final state
that consists of a high concentration of adopters.  In the converse case, the
stationary state is reached much more quickly, but the final fraction of
adopters is much lower and is severely limited by the significant densities of
Luddites and ignorants.

Since most models of innovation diffusion are formulated at mean-field level,
an important aspect of this work has also been to reveal the limitations of
the mean-field approximation. In particular, for Erd\H os-R\'enyi random
graphs with low mean degree and one-dimensional lattices, the mean-field
approximation proves inaccurate. This is due to the formation of Luddites 
which isolate domains of ignorants from the innovation, an effect particularly
apparently in one dimension.
It would be worthwhile to investigate the LISA model on
modular networks, where Luddism has the potential to block the spread of
innovation to entire communities.  In
addition to the work described in this paper, we also found that the
mean-field approximation proves better on two-dimensional lattices than on
one-dimensional lattices. 

In summary, the LISA model is a simple, but non-trivial, innovation diffusion
model that accounts for the possibility that the promotion of an innovation
may be tempered by the alienation of some individuals.  These in turn affect
the spread of the innovation. Interestingly, our model outlines two possible
marketing scenarios: If one is interested in reaching a high level of
adoption then this can only be achieved over long time scales, since the rate
of adoption must be low.  However, if the priority is to attain a finite
level of adoption as quickly as possible regardless of the alienation that
this may cause, then a high rate of adoption is preferable.

\section{Acknowledgements}

This work is supported by an EPSRC Industrial Case Studentship Grant
No. EP/L50550X/1.  SR is supported in part by NSF Grant No.\ DMR-1205797.
Partial funding from Bloom Agency in Leeds U.K. is also gratefully
acknowledged.


\begin{appendix}
\section{Analysis of one-dimensional dynamics}
\label{ap:1D}

In this appendix we describe the calculation of the final fractions of each
type of agent on one-dimensional lattices. These results are compared with
simulations in Fig.~\ref{fig:1dlattice_gamma} of Section~\ref{sec:1D}.

\subsection{Analysis of ignorant domains}

Initially, the nodes on the one-dimensional lattice are either ignorant, with
probability $I_0$, or susceptible, with probability $S_0=1-I_0$.  Thus the
initial configuration consists of connected domains of ignorant nodes
bordered by susceptibles.  Moreover, since ignorants can only become
susceptible if a neighbor is susceptible, domains of ignorants only evolve at
their ignorant-susceptible interfaces.  We will refer to these as ``active
interfaces''.  At an active interface one of three events can occur:
\begin{itemize}
\item The ignorant node becomes susceptible, thus reducing the domain length
  by one, with probability
\begin{equation*}
p_S= \frac{1/N}{1/N+r\gamma/2+\gamma}.
\end{equation*}

\item The ignorant node becomes a Luddite, thus reducing the length of the
  domain by one and causing the interface to become inactive, with
  probability
\begin{equation*}
p_L= \frac{r\gamma/2}{1/N+r\gamma/2+\gamma}.
\end{equation*}

\item The susceptible node becomes an adopter, thereby terminating the
  interface evolution, with probability
\begin{equation*}
p_A= \frac{\gamma}{1/N+r\gamma/2+\gamma}.
\end{equation*}
\end{itemize}

For an isolated ignorant node with two susceptible neighbors, these
probabilities respectively become
\begin{align*}
\hat{p}_S= &\frac{2/N}{2/N+r\gamma+\gamma},\\
\hat{p}_L= &\frac{r\gamma}{2/N+r\gamma+\gamma}, \\
\hat{p}_A= &\frac{\gamma}{2/N+r\gamma+\gamma}.
\end{align*}

Let $Q_n(m)$ be the probability that a domain of ignorants of initial
length $n$ with a \emph{single} ignorant-susceptible interface has a
\emph{final} length $n-m$, with $0\le m\le n$.   
We can determine $Q_n(m)$ as follows: If the final length of ignorants is
$n-m$, with $0<m<n$, then either $m$ ignorant nodes must become susceptible
before a susceptible node at the interface adopts, or $m-1$ ignorant nodes
must become susceptible before an ignorant node at the interface becomes a
Luddite. These events occur with probabilities $p_Ap_S^{m}$ and
$p_Lp_S^{m-1}$ respectively.  Using similar reasoning for the cases $m=0$ and
$m=n$, we thus find
\begin{equation}
Q_n(m)=\left\{
\begin{array}{crl}
p_A & \text{if} & m=0\\
p_Ap_S^{m}+p_Lp_S^{m-1} & \text{if} & 0<m<n\\
p_S^n+p_Lp_S^{n-1} & \text{if} & m=n
\end{array}
\right. .
\label{eq:Q}
\end{equation}
By summing over $m$, it can be shown that $Q_n(m)$ is normalized.

We now consider the case where a connected region of $n$ ignorant nodes
initially has two ignorant-susceptible interfaces. The probability $P_n(m)$
that a region of ignorants of initial length $n$ with \emph{two} active
interfaces has final length $n-m$ is given by the recursion relation
\begin{align}
P_n(m)&=Q_n(m)p_A+Q_{n-1}(m-1)p_L\nonumber\\
&\hskip 1cm +P_{n-1}(m-1)p_S,\label{eq:recur2}
\end{align}
where the terms $Q_n(m)$ are given by \eqref{eq:Q}.
Equation~\eqref{eq:recur2} captures the three possible events that can occur
at the interface.  If a susceptible node at the interface adopts, which
occurs with probability $p_A$, then the region of ignorants only has one
remaining active interface left and there will be $n-m$ remaining ignorants
with probability $Q_n(m)$, as given in \eqref{eq:Q}. If an ignorant node at the interface
becomes a Luddite, which occurs with probability $p_L$, then again the region
of ignorants will only have one active interface. Since there will be one
ignorant less the probability there will be $n-m$ remaining ignorants is
$Q_{n-1}(m-1)$.  Finally, if an ignorant node at the boundary becomes
susceptible, which occurs with probability $p_S$, then the probability that
there are $n-m$ ignorants remaining is the same as if we had started with $n-1$
ignorant nodes, i.e. $P_{n-1}(m-1)$.

To solve the recursion relation \eqref{eq:recur2} we need $P_n(0)$ and
$P_1(1)$.  The probability that a region of ignorants of initial length $n$
remains of length $n$ is given by
\begin{equation*}
P_n(0)=\left\{
\begin{array}{crl}
p_A\hat{p}_A & \text{if} & n=1\\
p_A^2 & \text{if} & n>1
\end{array}
\right. .
\label{eq:Pmm}
\end{equation*}
Also, the probability that a single ignorant node that initially has
two susceptible neighbors becomes a susceptible or Luddite is given
by
\begin{equation*}
P_1(1)=\hat{p}_A(p_L+p_S)+\hat{p}_L+\hat{p}_S.
\label{eq:P01}
\end{equation*}
Thus the solution to the recursion relation \eqref{eq:recur2} for
$0<m<n-1$ is given by
\begin{align*}
P_n(m)&=(m+1)p_A^2p_S^m+2mp_Ap_Lp_S^{m-1}\\
&+(m-1)p_L^2p_S^{m-2} .
\end{align*}
For $m=n-1$ we have
\begin{align*}
P_n(n-1)&=p_A\left[\hat{p}_A+(n-1)p_A\right]p_S^{n-1}\\
&+2(n-1)p_Ap_Lp_S^{n-2}+(n-2)p_L^2p_S^{n-3},
\end{align*}
and for $m=n$ we have
\begin{align*}
P_n(n)&=\left[\hat{p}_A(p_L+p_S)+\hat{p}_L+\hat{p}_S\right]p_S^{n-1}\\
&+(n-1)\left(p_Ap_S^n+2p_Lp_S^{n-1}+p_L^2p_S^{n-2}\right).
\end{align*}
Again it is possible to check, by summing \eqref{eq:recur2} over $m$
and solving the resulting recursion relation, that $P_n(m)$ is
normalized.

We can use $P_n(m)$ to calculate the expected final length of ignorant
domains $\langle x \rangle$. First note that since $I_0$ is the
initial probability of being ignorant, the probability that a domain
of ignorants initially has length $n>0$ is given by
$p_0(n)=I_0^{n-1}S_0$ for large $N$. Thus we find that
\begin{equation*}
\langle x\rangle
=\sum_{n=0}^Nnp_0(n)-\sum_{n=0}^Np_0(n)\sum_{l=0}^nlP_n(l).
\label{eq:meanx}
\end{equation*}
In principle, we may use the above to obtain an explicit expression for
$\langle x \rangle$.  In practice, however, we use the solutions to \eqref{eq:recur2} to
calculate $\langle x \rangle$ numerically.

\subsection{Calculation of population densities}
Initially, the mean number of ignorants is given by $I_0 N$ and so dividing
by the mean length of ignorant domains, $1/(1-I_0)$, yields the expected
number of ignorant domains, $(1-I_0)I_0N$.
Thus the final density of ignorants is $$I_\infty=(1-I_0)I_0
\langle x \rangle.$$

The probability that an ignorant domain survives is
$$q= 1 - \sum_{n=0}^{\infty}p_0(n)P_n(n).$$
Surviving ignorant domains have two interfaces, which are either
ignorant-adopter or ignorant-Luddite, with probabilities $p_A/(p_L + p_A)$
and $p_L/(p_L + p_A)$, respectively. Thus the expected number of Luddites at
the interfaces of non-vanishing ignorant domains is given by
\begin{equation}
\eta_{+}=\frac{2 p_L}{p_L + p_A} q(1-I_0)I_0 N.
\label{eq:edgeLuddites}
\end{equation}

It is also possible for Luddites to arise when a domain vanishes.  By
identifying the terms in $P_n(n)$ that result in Luddites, it is possible to
determine that the expected number of Luddites that arise when a domain of
initial size $n>1$ vanishes is given by
$$l_n=\left(\hat{p}_Ap_L+\hat{p}_L\right)p_S^{n-1}
+(n-1)\left(2p_Lp_S^{n-1}+p_L^2p_S^{n-2}\right)$$ and
$l_1=\hat{p}_Ap_L+\hat{p}_L$. Thus the expected number of Luddites
that arise from domains of ignorants that vanish is
\begin{equation}
\eta_0=(1-I_0)I_0N\sum_{n=0}^{\infty}p_0(n)l_n.
\label{eq:vanishLuddites}
\end{equation}
Summing Eqs.~\eqref{eq:edgeLuddites} and \eqref{eq:vanishLuddites} and
dividing by $N$ we arrive at the final density of Luddites
$$L_\infty = I_0(1-I_0) \left( \frac{2 p_L}{p_L + p_A} q + \sum_{n=0}^{\infty}p_0(n)l_n \right).$$
Since the dynamics cease when $S=0$, the number of adopters can be
found using the conservation law $A_\infty= 1 - L_\infty - I_\infty$.

\end{appendix}



\begin{thebibliography}{99}

\bibitem{R03} E. M. Rogers, \emph{Diffusion of Innovations} (Free Press, New
  York, 2003).
%
\bibitem{Kincaid04} D.~L. Kincaid, J. of Health Comm. {\bf 9}, 37 (2004).
%
\bibitem{R&G43} B.~Ryan and N.~Gross, Rural Sociology {\bf 8}, 15 (1943).
%
\bibitem{CKM}
  J.~Coleman, E.~Katz, and H.~Menzel, Sociometry {\bf 20}, 253 (1957).
%
\bibitem{1960s}
L.~A. Fourt and J.~W. Woodlock, Journal of Marketing {\bf 25} 31 (1960);
E. Mansfield, Econometrica {\bf 29}, 741 (1961).
%
\bibitem{B69} F.~M. Bass, Management Science {\bf 15}, 215 (1969).
%
\bibitem{B80} F.~M. Bass, J. Business {\bf 53}, S51 (1980).
%
\bibitem{M90}
V.~Mahajan, E.~Muller, and F.~M. Bass, Journal of Marketing {\bf 54}, 1 (1990).
%
\bibitem{B04} F.~M. Bass, Management Science {\bf 50}, 1833 (2004).
%
\bibitem{Hopp04}
W.~J. Hopp, Management Science {\bf 50}, 1763 (2004).
%
\bibitem{BMapplications}
M. G. Dekimpe, P. M. Parker and M. Sarvary, Technol. Forecast. Soc. Change {\bf 57}, 105 (1998);
S. Sundqvista, L. Franka and K. Puumalainen, J. Bus. Res. {\bf 58}, 107 (2005);
C. Michalakelis, D. Varoutas and T. Sphicopoulos, Telecommun. Policy {\bf 32}, 234 (2008):
J. Lim, C. Nam, S. Kim, H. Rhee, E. Lee and H. Lee, Telecommun. Policy {\bf 36}, 858 (2012);
J.~L. Toole, M.~Cha, and M.~C. Gonzalez, PLoS One {\bf 7}, e29528 (2012).
%
%
\bibitem{CEJOR2012}
E.~Kiesling et al., CEJOR {\bf 20}, 183 (2012).
%
\bibitem{PY09}
H.~Peyton Young, American Economic Review, {\bf 99}, 1899 (2009).
%
%
\bibitem{Sultan90}
F.~Sultan, J.~U. Farley, D.~R. Lehmann DR, Journal of Marketing {\bf 27} 70 (1990). 
%
\bibitem{BMparameters}
R.~M. Heeler, T.~P. Hustad, Management Science {\bf 26}, 1007 (1980);
C. Van den Bulte, G.~L. Lilien, Marketing Science {\bf 16}, 338 (1997);
R.~Kohli, D.~R. Lehmann, J. Pae, J.~Product Innovation Management {\bf 16}, 134 (1999).
%
\bibitem{Hohnisch08}
M~Hohnisch, S.~Pittnauer, and D.~Stauffer, Industrial and Corporate Change {\bf 17}, 1001 (2008).
%
%
\bibitem{C10} D. Centola, Science {\bf 329}, 1194 (2010).  
%
\bibitem{Cth}
 D. Centola, R. Wilker, and M. W. Macy, Am.\ J. Sociol. {\bf
    110}, 1009 (2005); D. Centola, V. M Eguiluz, and M. W. Macy, Physica A
  {\bf 374}, 449 (2007).
%
\bibitem{reinf} 
 P. S. Dodds and D. J. Watts, Phys.\ Rev.\ Lett.\ {\bf 92},
  218701 (2004); H. P. Young, Amer.\ Econ.\ Rev.\ {\bf 99}, 1899 (2009);
P.~L. Krapivsky, S.~Redner, and D. Volovik, J.~Stat.~Mech. {\bf P12003} (2011).
%
\bibitem{G00}
J. Goldenberg, B. Libai, S. Solomon, N. Jan N, and D. Stauffer, 
Physica A {\bf 284}, 335 (2000); D. Strang D and M.~W. Macy, American Journal of Sociology {\bf 107}, 147 (2001).
%
\bibitem{bandWagon}
P.~S. Tolbert and L.~G. Zucker, Admin.~Science Quarterly {\bf 28}, 22 (1983); 
E.~Abrahamson and L. Rosenkopf, Acad.~Management Review {\bf 18}, 487 (1993);
L. Rosenkopf and E.~Abrahamson, Comp. and Math. Organization Theory {\bf 5}, 361 (1999).
%

\bibitem{mobiles1}
http://www.pewinternet.org/data-trend/mobile/device-ownership/

\bibitem{mobiles2}
http://interphone.iarc.fr/UICC$\_$Report$\_$Final$\_$03102011.pdf

\bibitem{vaccine1}
R.~M. Wolfe and L.~K. Sharp, BMJ {\bf 325}, 430 (2002). 

\bibitem{vaccine2}
http://www.hscic.gov.uk/catalogue/PUB14949/nhs-immu-stat-eng-2013-14-rep.pdf

\bibitem{DOInets}
D.~J. Watts and P.~S. Dodds,  J Cons Res {\bf 34}, 441 (2007);
G. Kocsis and F. Kun, J.~Stat.~Mech.~P10014 (2008);
M.~Lin, N.~Li, Physica A {\bf 389}, 473 (2010);
 G. Pegoretti, F. Rentocchini, G.~V. Marzetti,  J.~Econ.~Interact.~Coord. {\bf 7}, 145 (2012);
N.~J. McCullen, A.~M. Rucklidge, C.~S.~E. Bale, T.~J. Foxon, and W.~F. Gale,  SIAM J. Applied Dynamical Systems {\bf 12}, 515 (2013);
S. Melnik, J.~A. Ward, J.~P. Gleeson, M.~A. Porter, Chaos, {\bf 23}, 013124 (2013);
K. Sznajd-Weron, J. Szwabinski, R. Weron, T. Weron, J.Stat.Mech. (2014) P03007;
P. Przybyla, K. Sznajd-Weron, R. Weron, Advances in Complex Systems 17 (2014) 1450004

\bibitem{resistance}
 S. Moldovan and J. Goldenberg, Technological Forecasting \& Social Change {\bf 71}
(2004) 425–442.

\bibitem{Luddites_hist}
http://www.nationalarchives.gov.uk/education/politics/g3/

\bibitem{Gillespie}
D.~T. Gillespie, J. Comput. Phys. {\bf 22}, 403 (1976).

\bibitem{Newman2010}
M. Newman, {\it Networks: An Introduction} (Oxford University Press, Oxford, 2010);
M.~O. Jackson, {\it Social and Economic Networks} (Princeton University Press, Princeton, 2010).

\bibitem{noise}
	 N. G. Van Kampen, {\it Stochastic Processes in Physics and Chemistry} (Elsevier, 2007);
	C. Gardiner, {\it Stochastic Methods} (Springer, 2010).
\bibitem{1D}  
{\it Nonequilibrium Statistical Mechanics in One Dimension}, edited by V. Privman (Cambridge University Press, New York, 1997);
S.~Redner, {\it A Guide to First-Passage Processes}
(Cambridge University Press, New York, 2001); P.~L. Krapivsky, S. Redner and E. Ben-Naim,
{\it A kinetic view of statistical physics} (Cambridge University Press, New York, 2010).


\end{thebibliography}
\end{document}